\documentclass[usenatbib]{basi}
\usepackage[T1]{fontenc}
\usepackage[british]{babel}
\usepackage[varg]{txfonts}
\usepackage{rotating}
\usepackage{dcolumn}
\begin{document}
\title[]{Analysis of the solar coronal green line profiles from eclipse observations}
\author[]
       {Maya Prabhakar$^1$\thanks{email: \texttt{mayap575@gmail.com}},
       K.P. Raju$^{1}$\thanks{email: \texttt{kpr@iiap.res.in}}
 and T. Chandrasekhar$^2$\thanks{email: \texttt{chandra@prl.res.in}}\\
       $^1$ Indian Institute of Astrophysics, Bangalore, India\\
       $^2$ Physical Research Laboratory, Ahmedabad, India}

\pubyear{2013}
\volume{00}
\status{submitted}


\maketitle
\label{firstpage}
\begin{abstract}
Analysis of the solar coronal green line profiles reveals information regarding
 the physical conditions of the solar corona like temperature, density, Doppler 
velocity, non-thermal velocity etc. It provides insights to the unresolved problems 
like the coronal heating and the acceleration of the solar winds. Recent studies have 
reported excess blueshifts in the coronal line profiles and are interpreted as due to
nanoflare heating, type II spicules and nascent solar wind flow. We have analyzed a 
time series of Fabry-Perot interferograms of the solar corona obtained during the total
solar eclipse of 2001 June 21 from Lusaka, Zambia. The spatial behavior of the coronal 
green line profiles were examined and variations in intensity, linewidth, Doppler velocity 
and line asymmetry were obtained. Several line profiles showed asymmetry indicating the 
presence of multicomponents. Such line profiles were fitted with double Gaussian curves. 
It has been found that 42\% of the line profiles were single components, 34\% were blueshifted, 
and 24\% were red shifted. The secondary component of a typical line profile 
with blue asymmetry is found to have a relative intensity about 0.26, Doppler velocity around -30 km/s 
and a halfwidth about 0.65$\AA$.
\end{abstract}


\section{Introduction}\label{s:intro}
The solar corona has very high 
temperatures of the order of millions of degrees. The basic problem posed by 
the discovery of the coronal temperature is to find the mechanism responsible 
for heating these layers by non-thermal energy transport which has remained an 
enigma since its discovery in 1940. Various mechanisms like 
magnetoacoustic waves, Alfven waves, magnetic reconnection, nanoflares, 
type-II spicules etc, constrained by velocity spectrum related to the 
magnetic field are attributed to cause the coronal heating.

Study of the emission coronal spectrum provides useful information regarding 
the physical conditions of the solar corona like temperature, density, Doppler 
velocity and non-thermal velocity. This may also provide insights to the 
unresolved mysteries like coronal heating and acceleration of solar winds. 
The emission coronal green line FeXIV 5302.86 $\AA$ is the strongest and widely 
studied line as its formation temperature (2 MK) is close to the average 
temperature of the inner solar corona. The presence of mass motions can change the shape 
of the line profiles. Raju (1998)  reports the interconnection between the presence 
of loops in the solar corona  and the structure of their line profiles.

Analysis of the coronal spectral lines was limited in the pre-Solar and Heliospheric 
Observatory (SOHO) era due to instrumental limitations (Domingo et al. 1995). The Coronal 
Diagnostic Spectrometer (CDS)  lacked a good spectral resolution  (Harrison et al. 1995) 
and Solar Ultraviolet Measurement of Emitted Radiation (SUMER) did not posses appreciable 
signal-to-noise ratio. Later the Extreme ultraviolet Imaging Spectrometer 
(EIS, Korendyke et al. 2006; Culhane et al. 2007) on board Hinode solar space observatory 
(Kosugi et al. 2007), in spite of having a low spectral resolution of 4000, enabled detailed 
study of the coronal line profiles due to good signal-to-noise ratio.

Many earlier coronal observations have reported small or no velocities and mass motions 
(Liebenberg, Bessey \& Watson 1975; Singh, Bappu \& Saxena 1982), while several others have 
reported large velocities and mass motions (Delone \& Makarova 1969; Desai, Chandrasekhar \& Angreji 1982; 
Raju et al. 1993). The SOHO and Transition Region and Coronal Explorer (TRACE) resolved this 
controversy by demonstrating that the solar atmosphere is much more variable and dynamic (Brekke 1999). 
Multiple flows were observed in active region by Kjeldseth-Moe et al. (1988, 1993).
De Pontieu (2009) finds that the non-thermal energy is insufficient to heat the coronal plasma 
to millions of degrees. Observations from Hinode and SOHO reveal significant coronal mass supply 
from the chromosphere showing ubiquitous upflows of velocities of the order of  50-100 km/s. 
These upflows are chromospheric jets named as type-II spicules. Data from EIS onboard 
Hinode reports strong blueshifts upto 50 km/s at the footpoints of the loops in the active regions 
(Peter 2010). Brekke (1999) on the other hand,  reports that the emission lines in the solar transition region are 
redshifted on an average.

Analysis of coronal green line profiles by Raju et al. (2011) shows excess blueshifts. 
Our study, a continuation of the latter involves the analysis of a series of Fabry-Perot interferograms 
of the total solar eclipse of June 21, 2001 obtained from Lusaka, Zambia. The magnitude 
of this eclipse was 1.0495. The data, analysis steps, results and conclusions are described 
in the following sections.

\section{Data and analysis}\label{s:data}
The instrumentation is similar to that used by Chandrasekhar et al. (1984). 
The free spectral range is 4.95 $\AA$ and the instrumental width is 0.2 $\AA$. The spectral 
resolution of the coronal green line is 26,000. The pixel size of the interferogram 
is 8 arc-sec. One of the interferogram we have analyzed is shown in Figure 1 (left).
\begin{figure}
\centerline{\includegraphics[scale=0.3]{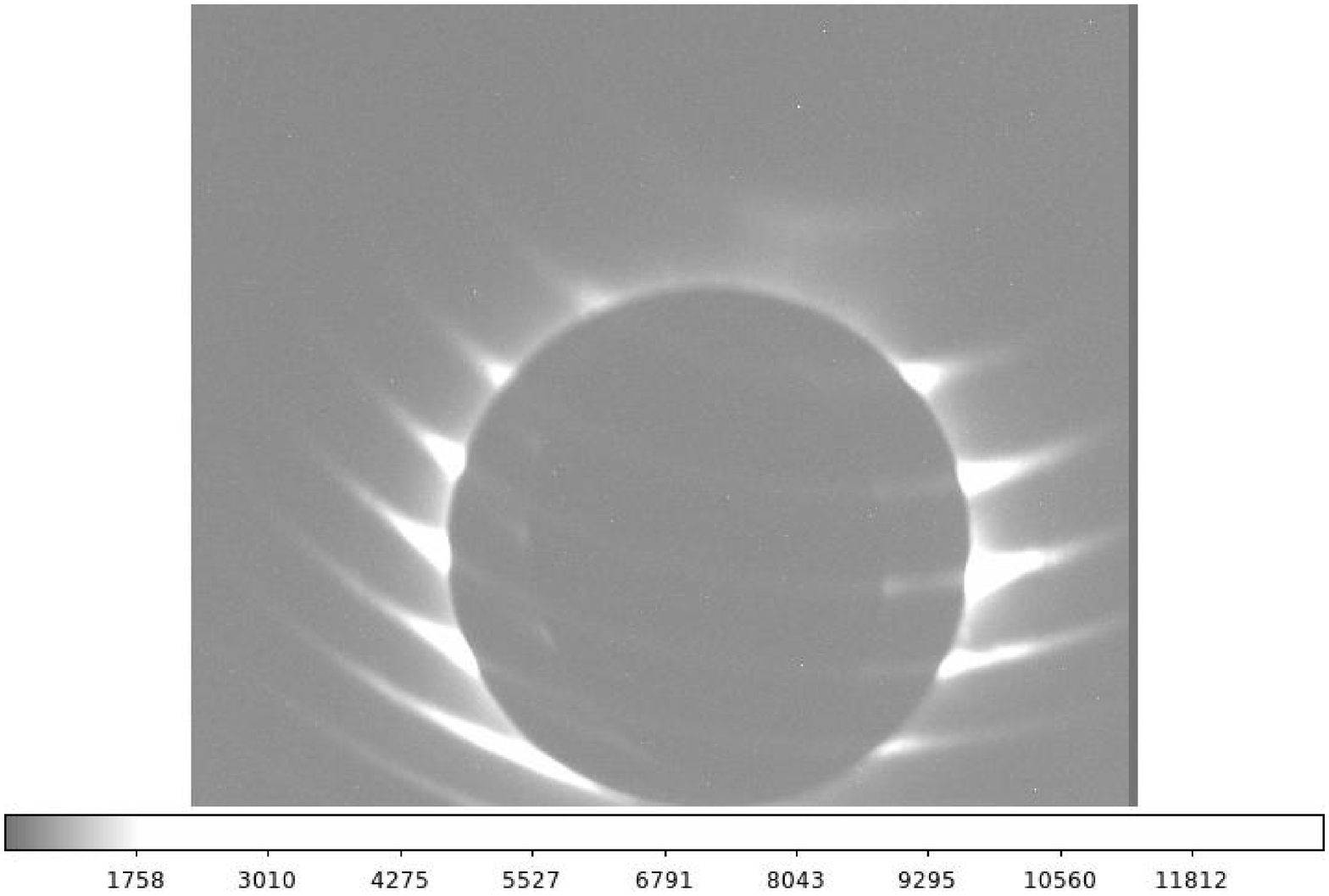}\includegraphics[scale=0.43]{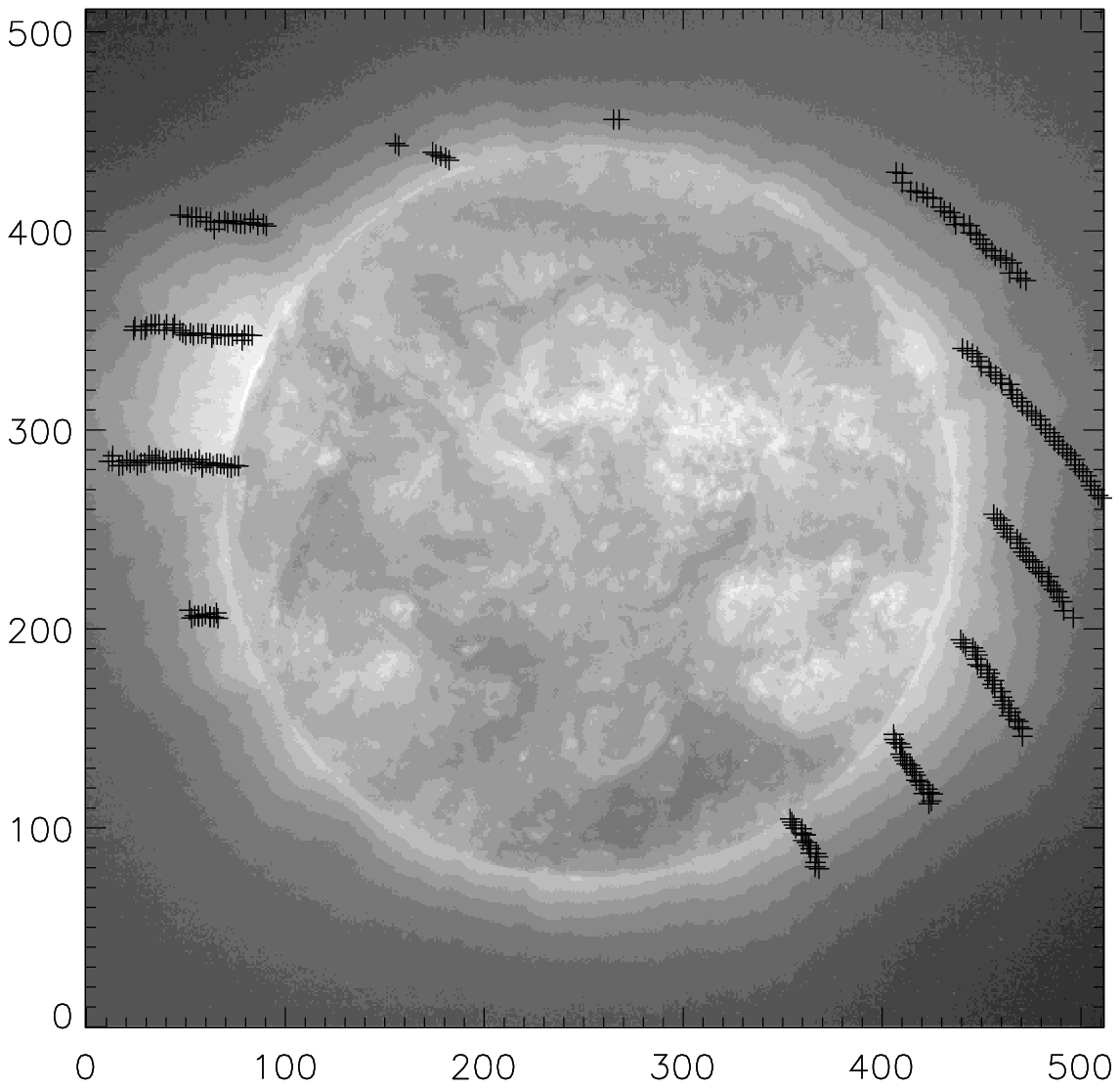}}
\caption{Left:A typical Fabry-Perot interferogram, right: Spatial locations of the line profiles plotted 
on the EIT image of the Sun}
\label{f:one}
\end{figure}
The analysis involves the following steps:
(1) Locating the fringe center and the solar center in the interferogram,
(2) Radial scans from the fringe center,
(3) Positional identification of the corona,
(4) Wavelength calibration,
(5) Continuum subtraction, and
(6) Gaussian fitting to the line profiles.
We obtained around 170 line profiles in the spectral 
range of 1.0-1.46 $R_{\odot}$ in a position angle coverage of $240^{\circ}$  
by radially scanning the interferograms. The line profiles were initially fitted with single Gaussian. 
The asymmetry in the line profiles is studied by determining the centroid, the mean 
position of the wavelength that divides the area of the line profile into two. 
Line profiles with multicomponents were fitted with double Gaussian curves to study their shifts.

\section{Results and discussions}\label{s:results}
The spatial locations of the analyzed line profiles are plotted on the Extreme 
ultraviolet Imaging Telescope (EIT) image of the Sun in Figure 1 (right), 
where north is up and west is to the right.
\begin{figure}
\centerline{\includegraphics[scale=0.4]{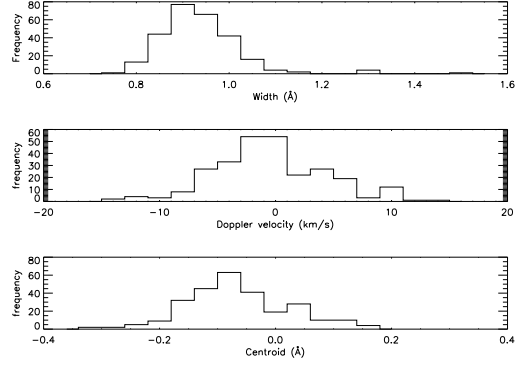}}
\caption{Histograms of width, Doppler velocity and centroid}
\label{f:two}
\end{figure}
Among the observed line profiles, 42\%
 were single components with negligible shifts, 34\% 
were blueshifted and 24\% were redshifted. 
The asymmetric line profiles being greater in number than the symmetric ones, 
in general indicate the presence of multicomponents in coronal green line profiles. 
The presence of blueshifts was first pointed out by Raju et al. (1993). 
Later the multicomponents were explained in terms of mass motions in the coronal loops (Raju 1999). 
Our results  agrees well with these observations. Histograms of width, Doppler velocity and centroid are
shown in Figure \ref{f:two}. It can clearly be seen that the histograms of Doppler velocity and 
centroid show excess blueshifts. The width peaks at 0.9 $\AA$, which when converted to temperature is 3.14 MK.
\begin{figure}
\centerline{\includegraphics[scale=0.4]{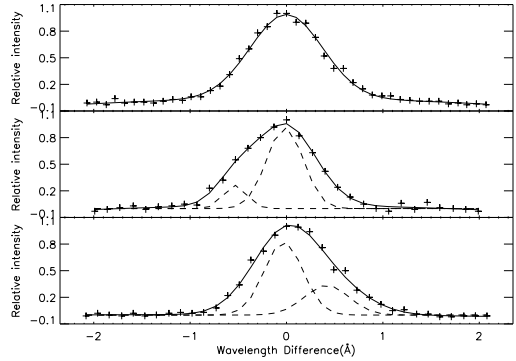}}
\caption{Gaussian fitting to the line profiles. Details are given in Table 1.}
\label{f:three}
\end{figure}
As the formation temperature of the line is 2 MK, this implies a non-thermal velocity of 20 km/s. 
The Gaussian fittings of the line profiles is shown in Figure \ref{f:three}. Those with single components 
were fitted with single Gaussian curves as shown in the first panel. Profiles showing asymmetry 
and larger Doppler velocity were fitted with double Gaussian curves. The second panel shows a profile 
with excess blueshift. The secondary component in it is found to have a relative intensity around 0.26, Doppler velocity around -30 km/s and a width around 0.65 $\AA$. Similarly, the third panel shows a profile with excess redshift. The red-shifted secondary component has a relative intensity around 0.33, Doppler velocity around -23 km/s and a width of 1.27 $\AA$. Details of the Gaussian fitting can be seen in Table 1.
\begin{table}
\begin{tabular}{|c|c|c|c|c|c|c|c|c|c|}
\hline
\multicolumn{1}{|c|}{}&\multicolumn{3}{|c|}{}&\multicolumn{3}{|c|}{}&\multicolumn{3}{|c|}{}\\
\multicolumn{1}{|c|}{}&\multicolumn{3}{|c|}{Single Gaussian}&\multicolumn{3}{|c|}{Blue}&\multicolumn{3}{|c|}{Red}\\\hline
No. & Int & Vel & Wid & Int & Vel & Wid & Int & Vel & Wid \\ \hline
1 & 0.95 & -0.08 & 1.24 & ~ & ~ & ~ & ~ & ~ & ~ \\ 
2 & 0.92 & -0.90 & 1.03 & 0.26 & -30.15 & 0.65 & ~ & ~ & ~ \\ 
3. & 0.83 & -1.69 & 1.10 & ~ & ~ & ~ & 0.33 & 23.59 & 1.27 \\
\hline
\end{tabular}
\caption{Details of Gaussian fitting. Row 1 gives the parameters of 
single Gaussian fitting. Row 2 gives the parameters of double Gaussian fitting
with a blue component. Row 3 gives the parameters of double Gaussian fitting
with a red component.}
\end{table}
The differential rotation of the Sun causes preferential blueshifts to the east limb and redshifts 
to the west limb. But, the velocity  which is just around 2 km/s is comparable to the error involved 
(Raju et al. 2011). However, recent observations from SOHO and Hinode shows predominant blueshifts 
in the EUV region (Hara et al. 2008; De Pontieu et al. 2009; Peter 2010) which have been explained 
due to type-II spicules and nascent solar wind flow (Tu et al. 2005; Tian et al. 2010). Possible 
explanations of the multicomponents are wave motions, mass motions in coronal loops, type II spicules 
or nascent solar wind flow.

\section{Conclusions}\label{s:conclusions}
A detailed analysis has shown that the coronal green line profiles, in general, contain multicomponents.
The excess blueshifts over redshifts in the line profiles agrees with the recent Hinode findings.
The observed non-thermal velocities agrees well with the earlier reported results. The secondary component
of a typical line profile with blue asymmetry is found to have
a relative intensity about 0.26, Doppler velocity around -30 km/s and a halfwidth about 0.65$\AA$.
The causes for the line profile asymmetry is expected to be further probed from Hinode observations.

\label{lastpage}

\end{document}